\documentstyle[sprocl]{article}

\def\be{\begin{equation}}
\def\ee{\end{equation}}
\def\bea{\begin{eqnarray}}
\def\eea{\end{eqnarray}}
\def\lqcd{\Lambda_{\rm QCD}}
\def\mev{\,{\rm MeV}}
\def\gev{\,{\rm GeV}}
\def\vcb{|V_{cb}|}
\def\case#1#2{\textstyle{#1\over#2}}
\def\d{{\rm d}}

\begin{document}

\begin{flushright}
JHU--TIPAC--96014\\
hep-ph/9609380\\
September, 1996
\end{flushright}

\vspace{1cm}

\title{THE MANY USES OF EXCITED HEAVY HADRONS\footnote{To appear in the
Proceedings of the Twentieth Johns Hopkins Workshop on Current Problems in
Particle Theory, ``Nonperturbative Particle Theory and Experimental Tests'',
Heidelberg, Germany, June 27--29, 1996.}}

\author{ADAM F.~FALK}

\address{Department of Physics and Astronomy\\
The Johns Hopkins University\\
3400 North Charles Street\\
Baltimore, Maryland 21218 U.S.A.}

\maketitle\abstracts{I discuss a variety of issues in the physics of
excited bottom and charmed hadrons.  Recent developments in
spectroscopy, strong decays, and production in fragmentation and
weak decays are reviewed.}

\section{Introduction} The experimental and theoretical study of
heavy hadrons typically focuses on the ground state $D^{(*)}$ and
$B^{(*)}$ mesons, and on the lightest baryons, $\Lambda_c$ and
$\Lambda_b$.  This is hardly surprising, since these states are the
most copiously produced, and the most long-lived, making detailed
experiments possible.  Nonetheless, these are but the lightest states
in a tower of excitations.  These excited states are also
interesting, for a variety of reasons.  First, one can use them as a
laboratory to study the heavy quark (and light flavor) symmetries
which are crucial to much of heavy hadron phenomenology.  Second,
there are new applications of heavy quark symmetry, leading to new
questions about QCD, which only arise in the study of the more
complicated spin structure of excited heavy hadrons.  Third, certain
experiments involving these excitations yield new information which
is directly applicable to the physics of the ground state
heavy hadrons.

In this talk, I will briefly survey a variety of such issues.  I
begin with a review of hadron dynamics in the heavy quark
limit,\footnote{The reader who desires a more extensive introduction to
Heavy Quark Effective Theory, with thorough references to the
original literature, may consult a number of excellent reviews.~\cite{reviews}}
and of the simple
spectroscopic predictions which follow from it.  Certain of these
predictions are currently not well satisfied, casting doubt either on
the data or on its interpretation.  I will then turn to the strong
decays of heavy mesons and show that a consistent inclusion of
subleading effects can resolve an otherwise puzzling discrepancy.  The
third topic will be the production of excited heavy hadrons in
fragmentation processes, and the fourth will be the production of
heavy hadrons in semileptonic decays.  In the latter case, we will see
that it is possible to extract information from such processes which
is useful for improving the extraction of the CKM matrix element
$\vcb$ from semileptonic $B$ decays.

In addressing these four topics, I do not pretend to review the
entire field of excited bottom and charmed hadrons.  Rather, I hope
to illustrate the rich interplay between theory and experiment which
these recent developments make possible.

\section{The Heavy Quark Limit}\label{HQL}

Consider a hadron containing a single heavy quark $Q$, where by
``heavy'' we  mean that its mass satisfies the condition
$m_Q\gg\lqcd\sim500\mev$.  Ultimately, of course, we will apply
this analysis to physical charm and bottom quarks, with
$m_c\approx1.5\mev$ and $m_b\approx4.8\mev$, which may not be well
into this asymptotic regime.  Hence, at times it will be important
to include corrections which are subleading in an expansion in
$\lqcd/m_Q$.  For now, however, let us assume that we are in a regime
where this ``heavy quark limit'' applies.

A heavy hadron is a bound state consisting of the heavy quark and
many  light degrees of freedom.  These light degrees of freedom
include valence quarks and antiquarks, sea quarks and antiquarks,
and gluons, in a complex configuration determined by
nonperturbative strong interactions.  These
interactions are characterized by the dimensional scale $\lqcd$,
the scale at which the strong coupling $\alpha_s$ becomes of order 1;
in particular, $\lqcd$ is the typical energy associated with
the four-momenta carried by the light degrees of freedom. Hence it
is also the typical energy of quanta exchanged with the
heavy quark in the bound state.  Since $m_Q\gg\lqcd$, the heavy
quark does not recoil upon exchanging such quanta with the light
degrees of freedom.  This is the simple physical content of the heavy
quark limit:  {\it $Q$ acts as a static source of chromoelectric
field, so far as the light degrees of freedom are concerned.}
In a more covariant language, the four-velocity $v^\mu$ of
$Q$ is unchanged by the strong interactions.  Because the heavy quark does not
recoil from its interactions with
the light degrees of freedom, they are insensitive to its mass, so
long as $m_Q\gg\lqcd$.  This is analogous to the
statement in quantum electrodynamics that the electronic wave
function is the same in hydrogen and deuterium.

There is also a condition on the spin of the heavy quark, which
couples to the light degrees of freedom primarily through the
chromomagnetic interaction.  Since the
chromomagnetic moment of $Q$ is given by $g\hbar/2m_Q$, this
interaction also vanishes in the heavy quark limit.  Not only is
the velocity of the heavy quark unchanged by soft QCD, so is the
orientation of its spin.
Hence, if the light degrees of freedom have nonzero angular momentum
$J_\ell$, then the states with total $J=J_\ell+{1\over2}$ and
$J=J_\ell-{1\over2}$ are degenerate. This is analogous to the
statement in quantum electrodynamics that the
hyperfine splitting in hydrogen is much smaller than the electronic
excitation energies.  Thus we have new {\it symmetries\/} of the
spectrum of QCD in the heavy quark limit.~\cite{IW}  These lead to new ``good''
quantum numbers, the excitation energy and the total angular momentum
of the light degrees of freedom, which can be sensibly defined only in
this limit.

If we have $N_h$ heavy quarks, $Q_1\ldots Q_{N_h}$, then the heavy
quark symmetry group is $SU(2N_h)$.  These symmetries yield
relations between the properties of hadrons containing a single
heavy quark, including masses, partial decay widths, and weak form
factors.  These relations can often be sharpened by the systematic
inclusion of effects which are subleading in the $1/m_Q$ expansion.

\section{Spectroscopy}\label{SPEC}

The simplest heavy quark relations are those for the
spectroscopy of states containing a single heavy quark.  Heavy
hadron spectroscopy differs from that for hadrons containing only
light quarks because we may specify separately the spin quantum
numbers of the light degrees of freedom and of the heavy quark.
The constituent quark model can serve as a useful guide for
enumerating these states.  Of course, we should not take this model
too seriously, as it has certain unphysical features, such as
drawing an additional distinction between spin and orbital angular
momentum of the valence quarks, and including explicitly neither sea
quarks nor gluons.  So remember in what follows that any mention of
constituent quarks is purely for the purpose of counting quantum
numbers.

\subsection{Heavy mesons}

In the constituent quark model, a heavy meson consists of a heavy
quark and a light antiquark, each with spin ${1\over2}$, in a
wavefunction with a given excitation energy and a given orbital
angular momentum.  There is no natural zero-point with respect to
which to define energies in this confined system, but differences
between energy levels $E_\ell$ of the light antiquark are
well defined.  The antiquark can have any integral orbital angular
momentum, $L=0,1,2,\ldots$, with parity $(-1)^L$.  Combined with
the intrinsic spin-parity $S_\ell^P={1\over2}^-$ of the antiquark, we
find states with total spin-parity
\be
  J_\ell^P=\case12^\pm,\case32^\pm,\case52^\pm,\ldots\,.
\ee
This is then added to the spin parity $S_Q^P={1\over2}^+$ of the
heavy quark, to yield states with total angular momentum
\be
  J^P= 0^\pm,1^\pm,2^\pm,\ldots\,.
\ee
In the limit $m_Q\to\infty$, the two states with a given $J_\ell^P$
are degenerate.

As an example, let us consider the charmed mesons.
Our quark model intuition tells us correctly that the ground state
light degrees of freedom have the quantum numbers of a light antiquark
in an $s$ wave, so $J_\ell^P={1\over2}^-$ and the two degenerate
states have $J^P=0^-$ and $1^-$.  This is indeed what is observed:
a $0^-$ $D$ meson with mass approximately $1870\mev$, and a slightly
heavier $1^-$ $D^*$ at about $2010\mev$.  (I will keep to
approximate masses for now, as I do not want to concern myself with
small isospin splittings which complicate the situation in an
unimportant way.)  The nonzero splitting between the $D$ and the
$D^*$ is an effect of order $1/m_c$; this splitting scales as
$\lqcd^2/m_c$ in the heavy quark expansion.

As the next excitation, we might expect to
find the antiquark in a $p$ wave.  With the antiquark spin, we find
light degrees of freedom with $J_\ell^P={1\over2}^+$ and
$J_\ell^P={3\over2}^+$, each leading to an (almost) degenerate
doublet of states.  The doublet with $J^P=0^+$ and $1^+$
has not been observed, presumably because it is very broad (see
Section~\ref{STRONG}).  The other doublet, with $J_\ell^P={3\over2}^+$,
consists of the
$D_1(2420)$ and $D_2^*(2460)$.  The splitting between the $D_1$ and
the $D_2^*$, again a $1/m_c$ effect, is not related to the $D-D^*$
splitting.

The heavy quark symmetries imply relations between the spectra of
the bottom and charmed meson systems.  Because the mass of a heavy
hadron can be decomposed in the form $M_H=m_Q+E_\ell$, the entire
spectrum of bottom mesons can be determined from the charmed mesons
once the quark mass difference $m_B-m_c$ is known.\footnote{The
difficult question of how properly to define heavy quark masses is
not really relevant to heavy hadron spectroscopy.  For now, it is
best to take $m_Q$ to denote the pole mass at some fixed order in
QCD perturbation theory.}  This difference can be found, for
example, from the ground state mesons.  Taking a spin average to
eliminate the hyperfine energy,
\be
  \overline D = \case14(D+3D^*)\,,\qquad
  \overline B =\case14(B+3B^*)\,,
\ee
and letting the states stand for their masses, we
find
\be
  \overline B-\overline D=m_b-m_c\approx3.34\mev\,.
\ee
One then finds relations for the excited states, such as
\be
  \overline B_1-\overline D_1=\overline B-\overline D\,,
\ee
where $\overline D_1={1\over8}(3D_1+5D_2^*)$ is the appropriate spin average.
Including strange quarks, one finds
similar relations, such as $\overline B_s-\overline D_s=\overline
B-\overline D$.  There are also relations which exploit the known
scaling of the hyperfine splitting in the heavy quark limit.  Since
$D^*-D\sim\lqcd^2/m_c$, we find
\bea
  (B^*)^2-B^2&=&(D^*)^2-D^2\,,\\
  (B_{s2}^*)^2-B_{s1}^2&=&(D_{s2}^*)^2-D_{s1}^2\,,
\eea
and so on.

\begin{table}
  \caption{The observed charmed and bottom mesons.}
  \vspace{0.4cm}
  \centerline{
  \begin{tabular}{|c|ccccccc|}
  \hline
     \multicolumn{2}{|c}{Spin}&\multicolumn{3}{c}{$D$ system~~\cite{CLEO,PDG}}
           &\multicolumn{3}{c|}{$B$ system~~\cite{OPAL,PDG}}\\
     \hline
     $J_\ell^P$&$J^P$&state&$M$ (MeV)
     &$\Gamma$ (MeV)&state&$M$ (MeV)&$\Gamma$ (MeV)\\
     \hline
     ${1\over2}^-$&$0^-$&$D^0$&1865&$\tau=0.42\,
          $ps&$B^0$&5279&$\tau=1.50\,$ps\\
     &&$D^\pm$&1869&$\tau=1.06\,$ps&$B^\pm$&5279&$\tau=1.54\,$ps\\
     &&$D_s$&1969&$\tau=0.47\,$ps&$B_s$&5375&$\tau=1.34\,$ps\\
     \cline{2-8}
     &$1^-$&$D^{*0}$&2007&$<2.1$&$B^*$&5325&\\
     &&$D^{*\pm}$&2010&$<0.13$&&&\\
     &&$D_s^*$&2110&$<4.5$&$B_s^*$&&\\
     \hline
     ${1\over2}^+$&$0^+$&$D_0^*$&&&$B_0^*$&&\\
     \cline{2-8}
     &$1^+$&$D_1'$&&&$B_1'$&&\\
     \hline
     ${3\over2}^+$&$1^+$&$D_1^0$&$2421\pm3$&$20\pm7$&$B_1$&5725&20\\
     &&$D_1^\pm$&$2425\pm3$&$26\pm9$&&&\\
     &&$D_{s1}$&2535&$<2.3$&$B_{s1}$&5874&1\\
     \cline{2-8}
     &$2^+$&$D_2^{*0}$&$2465\pm4$&$28\pm10$&$B_2^*$&5737&25\\\
     &&$D_2^{*\pm}$&$2463\pm4$&$27\pm12$&&&\\
     &&$D_{s2}^*$&$2573\pm2$&$16\pm6$&$B_{s2}^*$&5886&1\\
  \hline
  \end{tabular}}
  \label{mesontable}
\end{table}

The charmed and bottom mesons which have been identified are listed
in Table~\ref{mesontable}, along with their widths (which will be
of interest in Section~\ref{STRONG}).  Given the measured properties of
the charmed mesons, we can make a set of predictions for the
bottom system,
\bea
  B^*-B=&\!\!\!\!\!\!52\mev&\qquad(46\mev)\\
  \overline B_1 =&5789\mev&\qquad(5733\mev)\\
  B_s^*-B_s=&\!\!\!\!\!\!14\mev&\qquad(12\mev)\\
  \overline B_{s1}=&5894\mev&\qquad(5882\mev)\\
  B^*_{s2}-B_{s1}=&\!\!\!\!\!\!13\mev&\qquad(12\mev)\,.
\eea
The experimental values are given in parentheses.  We can
estimate the accuracy with which we expect these predictions to hold by
considering the size of the largest omitted term in the expansion
in $1/m_B$ and $1/m_c$.  For relations between spin-averaged
quantities, this is
\be
  \lqcd^2\left({1\over2m_c}-{1\over2m_b}\right)\sim50\mev\,,
\ee
while for relations involving hyperfine splittings, we have
\be
  \lqcd^3\left({1\over4m_c^2}-{1\over4m_b^2}\right)\sim5\mev\,.
\ee
These estimates are confirmed by the results given above.

The relations we have derived here follow rigorously from QCD in
the heavy quark limit, $m_Q\to\infty$.  Of course, they also arise
in phenomenological models of hadrons, such as the nonrelativistic
constituent quark model.  In fact, an important test of any such
model is that it have the correct heavy quark limit.  Since the
constituent quark model has this property, it reproduces these
predictions as well.  However, unlike the heavy quark limit, the
constituent quark model is not in any sense a controlled {\it
approximation\/} to QCD, and it is impossible to estimate the error
in a quark model prediction in any meaningful way.

One intriguing feature of the quark model is that it makes accurate
predictions for many light hadrons, too.  It is not clear whether
these successes have a clear explanation, or even a single one.
Perhaps, at some length scales, nonrelativistic constituent quarks
really are appropriate degrees of freedom.  Perhaps its success
lies in its closeness to the large $N_c$ limit of QCD,~\cite{Manohar} in
which quark pair production is also suppressed.  Whatever the
proper explanation, it is important to keep in mind that relations
which follow solely from the quark model do not have the same status as
those that follow from real {\it symmetries\/} of QCD, such as
heavy quark symmetry or light flavor $SU(3)$.

\subsection{Heavy baryons}

Because heavy baryons contain two light quarks, their flavor
symmetries are more interesting than those of the heavy mesons;
however, because they are more difficult to produce, less is known
experimentally about the spectrum of heavy baryon excitations.  For
simplicity, let us restrict ourselves to states in which the light
quarks have no orbital angular momentum.  Then, given two quarks
each with spin ${1\over2}$, the light degrees of freedom can be in
an antisymmetric state of total angular momentum $J_\ell^P=0^+$ or a
symmetric state with  $J_\ell^P=1^+$.  By the Pauli exclusion
principle, if neither light quark is a strange quark then the
spin and isospin are the same.  The exclusion principle also
prohibits a $J_\ell^P=0^+$ state with two strange quarks.

\begin{table}
  \caption{The lowest lying charmed baryons  Isospin is denoted by
$I$, strangeness by $S$.}
  \vspace{0.4cm}
  \centerline{
  \begin{tabular}{|l|l|llllr|l|}
  \hline
  Name&$J^P$&$s_\ell$&$L_\ell$&$J^P_\ell$&$I$&$S$&Decay\\
  \hline
  $\Lambda_c$&$\case12^+$&0&0&$0^+$&0&0&weak\\
  $\Sigma_c$&$\case12^+$&1&0&$1^+$&1&0&
  $\Lambda_c\pi$, $\Lambda_c\gamma$, weak\\
  $\Sigma^*_c$&$\case32^+$&1&0&$1^+$&1&0&$\Lambda_c\pi$\\
  $\Xi_c$&$\case12^+$&0&0&$0^+$&$\case12$&$-1$&weak\\
  $\Xi'_c$&$\case12^+$&1&0&$1^+$&$\case12$&$-1$&$\Xi_c\gamma$, $\Xi_c\pi$\\
  $\Xi^*_c$&$\case32^+$&1&0&$1^+$&$\case12$&$-1$&$\Xi_c\pi$\\
  $\Omega_c$&$\case12^+$&1&0&$1^+$&0&$-2$&weak\\
  $\Omega^*_c$&$\case32^+$&1&0&$1^+$&0&$-2$&$\Omega_c\gamma$\\
  \hline
  \end{tabular}}
  \label{baryontable}
\end{table}

When the spin of the heavy quark is included, the $J_\ell^P=0^+$
state becomes a baryon with spin-parity $J^P={1\over2}^+$, while the
$J_\ell^P=1^+$ state becomes a doublet of baryons with
$J^P={1\over2}^+$ and $J^P={3\over2}^+$.  The quantum numbers of
the charmed baryons are listed in Table~\ref{baryontable}, along
with their expected decays.  Note that the dominant decay channels
of the higher mass $J^P={1\over2}^+$ states $\Sigma_c$ and $\Xi_c'$
are determined by the available phase space.  If emission of a pion
is possible, then they will decay strongly; if not, then they will decay
weakly or electromagnetically, depending on their charge.

Again, there are heavy quark symmetry relations between the bottom
and charmed systems.  The hyperfine interaction between the heavy
quark and the $J_\ell=1$ light degrees of freedom is removed by the
spin averages
\bea
  \overline\Sigma_c&=&\case13\left(\Sigma_c+2\Sigma_c^*\right)\\
  \overline\Xi_c&=&\case13\left(\Xi_c'+2\Xi_c^*\right)\\
  \overline\Omega_c&=&\case13\left(\Omega_c+2\Omega_c^*\right)\,.
\eea
Then we find heavy quark relations of the form
\bea
  \Lambda_b-\Lambda_c&=&\overline B-\overline D\label{hqrel1}\\
  \overline\Sigma_b-\Lambda_b&=&\overline\Sigma_c-\Lambda_c\label{hqrel2}\\
  {\Sigma_b^*-\Sigma_b\over\Sigma_c^*-\Sigma_c}&=&
  {B^*-B\over D^*-D}\,.\label{hqrel3}
\eea
We can also use light flavor $SU(3)$ symmetry to relate the
nonstrange charmed baryons to the charmed baryons with strange
quarks.  There are three relations which include corrections of order
$m_s$,~\cite{Savage}
\bea
  \Xi_c'&=&\case12\left(\Sigma_c+\Omega_c\right)\label{su3rel1}\\
  \Xi_c^*&=&\case12\left(\Sigma^*_c+\Omega^*_c\right)\label{su3rel2}\\
  \Sigma_c^*-\Sigma_c&=&\Xi_c^*-\Xi_c'\,.\label{su3rel3}
\eea

There is another relation in which corrections of order $m_s$ are
{\it not\/} systematically included,
\be
  \overline\Sigma_c-\Lambda_c=\overline\Xi_c'-\Xi_c\,;\label{su3rel4}
\ee
however, since the analogous relation in the charmed meson system,
\be
  \overline D_{s1}-\overline D_s=\overline D_1 -\overline D\,,
\ee
works
to within a few MeV, we will use this one as well.

The lightest observed heavy baryons are listed in
Table~\ref{baryondata}, along with their masses and the decay channels
in which they have been identified. I identify the observed states by
provisional names, while in the penultimate column I give the
conventional assignment of quantum numbers to these states.  These
assignments are motivated primarily by the quark model.

Let us compare the predictions of heavy quark and flavor $SU(3)$
symmetry to these experimental results.~\cite{Falk}  The heavy quark
constraints~(\ref{hqrel1}) and~(\ref{hqrel2}) are both satisfied to
within $10\,$MeV.  However, the hyperfine relation (\ref{hqrel3}) is
badly violated.  One finds
$(\Sigma_b^*-\Sigma_b)/(\Sigma_c^*-\Sigma_c)\approx0.84\pm0.21$, too
large by more than a factor of two!  (I have ignored the correlation
between the errors on the masses of the $\Sigma_b$ and the $\Sigma^*_b$,
thereby overestimating the total uncertainty.)  Clearly, if these data are
correct then there is a serious crisis for the application of heavy
quark symmetry to the charm and bottom baryons.  On the other hand,
this crisis rests {\it entirely} on the reliability of the DELPHI
measurement~\cite{DELPHI} of these states.

The situation is somewhat better for the $SU(3)$ relations, although
not perfect.  The first equal spacing rule (\ref{su3rel1}), yields the
prediction $\Xi'_c=2577\,$MeV, somewhat large but probably within the
experimental error.  The second rule (\ref{su3rel2}) cannot be tested,
as the $\Omega^*_c$ state has not yet been found.  The third
rule~(\ref{su3rel3}) yields the prediction $\Xi'_c=2578\,$MeV, again,
reasonably consistent with experiment.  (In fact the precise agreement
of these two sum rules might lead one to expect that, when confirmed,
the mass of the $\Xi_c'$ will be somewhat higher than its present
central value.)  By contrast, the final $SU(3)$
relation~(\ref{su3rel4}) fails by approximately 60~MeV, almost an
order of magnitude worse than for the charmed mesons.  However, this
relation is not on the same footing as the others, so its failure is
not as significant as that of the heavy quark relation (\ref{hqrel3}).

\begin{table}
  \caption{The observed heavy baryon states, with their conventional
and alternative identities.  Isospin multiplets have been averaged
over.  Statistical and systematic errors have, for simplicity, been
added in quadrature.  The approximate masses of the proposed new
states are given in parentheses.}
  \vspace{0.4cm}
  \centerline{
  \begin{tabular}{|l|lll|l|l|}
  \hline
  State&Mass (MeV)&Ref.&Decay&Conventional&Alternative \\
  \hline
  $\Lambda_c$&$2285\pm1$&~\cite{PDG}&weak&$\Lambda_c$&$\Lambda_c$\\
  &(2380)&&weak&absent&$\Sigma_c^{0,++}$\\
  &(2380)&&$\Lambda_c+\gamma$&absent&$\Sigma_c^+$\\
  $\Sigma_{c1}$&$2453\pm1$&~\cite{PDG}&$\Lambda_c+\pi$&$\Sigma_c$&
    $\Sigma^*_c$\\
  $\Sigma_{c2}$&$2519\pm2$&~\cite{CLEO96}&$\Lambda_c+\pi$&
    $\Sigma^*_c$&?\\
  $\Xi_c$&$2468\pm2$&~\cite{PDG}&weak&$\Xi_c$&$\Xi_c$\\
  $\Xi_{c1}$&$2563\pm15$\ (?)&~\cite{WA89}
    &$\Xi_c+\gamma$&$\Xi'_c$&$\Xi'_c$\\
  $\Xi_{c2}$&$2644\pm2$&~\cite{CLEO95}&$\Xi_c+\pi$&$\Xi^*_c$&$\Xi^*_c$\\
  $\Omega_c$&$2700\pm3$&~\cite{E687}&weak&$\Omega_c$&$\Omega_c$\\
  $\Omega_c^*$&not yet seen&&&&\\
  \hline
  $\Lambda_b$&$5623\pm6$&~\cite{PDG,CDF96}&weak&$\Lambda_b$&$\Lambda_b$\\
  &(5760)&&weak&absent&$\Sigma_b^\pm$\\
  &(5760)&&$\Lambda_b+\gamma$&absent&$\Sigma_b^0$\\
  $\Sigma_{b1}$&$5796\pm14$&~\cite{DELPHI}&$\Lambda_b+\pi$&
    $\Sigma_b$&$\Sigma^*_b$\\
  $\Sigma_{b2}$&$5852\pm8$&~\cite{DELPHI}&$\Lambda_b+\pi$&
    $\Sigma^*_b$&?\\
  \hline
  \end{tabular}}
   \label{baryondata}
\end{table}
What is going on here?  One possibility is that the heavy quark
relations are simply no good for the spectroscopy of charmed baryons.
Of course, we would like to avoid this glum conclusion, because it
would call into question other applications of heavy quark symmetry to
charmed hadrons, such as the treatment of exclusive semileptonic $B$
decays used to extract $\vcb$.  Another possibility is that
the data are not correct.  This may not be unlikely, particularly as the
discrepancy rests primarily on the single DELPHI measurement.
However, let us look for an alternative resolution, in which we take
the reported data seriously, within their reported errors.   As the data change
in the future, so perhaps will the motivation for such an alternative.

Let us, then, reinterpret the data under the constraint that the heavy
quark and $SU(3)$ symmetries be imposed explicitly, including the
dubious relation~(\ref{su3rel4}).~\cite{Falk}  Then if we identify the
observed
$\Xi_{c1}$ with the $\Xi'_c$ state, the $SU(3)$ relations lead to
the prediction $\Sigma_c=2380\mev$.  If this is true, then it cannot
be correct to identify the $\Sigma_c$ with the observed $\Sigma_{c1}$;
rather, the $\Sigma_c$ would correspond to a state below threshold for
the decay $\Sigma_c\to\Lambda_c+\pi$, which is yet to be seen.
The observed $\Sigma_{c1}$ must then be the $\Sigma_c^*$, while the
observed $\Sigma_{c2}$ is some more highly excited baryon, perhaps an
orbital excitation.  The new assignments are given in the final column
of Table~\ref{baryondata}.

A similar reassignment must be applied to the bottom baryons as well.
The $\Sigma_b$ is now assumed to be below $\Lambda_b+\pi$ threshold,
while the $\Sigma_{b1}$ is identified as the $\Sigma_b^*$.  Then the
poorly behaved symmetry predictions improve remarkably.  For example,
let us take the masses of the new states to be $\Sigma_c=2380\,$MeV
and $\Sigma_b=5760\,$MeV.  Then the hyperfine splitting ratio
(\ref{hqrel3}) improves to
$(\Sigma_b^*-\Sigma_b)/(\Sigma_c^*-\Sigma_c)=0.49$, and the $SU(3)$
relation (\ref{su3rel4}) between the $s_\ell=0$ and $s_\ell=1$ states
is satisfied to within $5\,$MeV.  The heavy quark
relation~(\ref{hqrel1}) is unaffected, while the
constraint~(\ref{hqrel2}) for the $\overline\Sigma_Q$ excitation
energy is satisfied to within $20\,$MeV, which is quite reasonable.
Only the $SU(3)$ equal spacing rules~(\ref{su3rel1})
and~(\ref{su3rel3}) suffer from the change.  The former relation now
fails by $23\,$MeV.  The latter now fails by $8\,$MeV, but the
discrepancies are in {\it opposite\/} directions, and the two
relations cannot be satisfied simultaneously by shifting the mass of
the $\Xi'_c$.  With these new assignments, intrinsic $SU(3)$ violating
corrections of the order of $15\,$MeV seem to be unavoidable.  In this
context, a confirmation of the $\Xi_c'$ state is very important.  If
the mass were to be remeasured to be approximately 2578~MeV, then
$SU(3)$ violation under the conventional assignments would be
extremely small and we might be more disinclined to relinquish them.

Still, with respect to the symmetry predictions as a whole, the new
scenario is quite an improvement over the old.  The heavy quark and
$SU(3)$ flavor symmetries have been resurrected.  We can improve the
agreement further if we allow the measured masses to vary within their
reported $1\sigma$ errors.  One set of allowed masses is
$\Sigma_c=2375\,$MeV, $\Sigma^*_c=2453\,$MeV, $\Xi'_c = 2553\,$MeV,
$\Xi^*_c=2644\,$MeV, $\Sigma_b=5760\,$MeV, and
$\Sigma^*_b=5790\,$MeV.  For this choice, the $SU(3)$ relations
(\ref{su3rel1}), (\ref{su3rel3}) and (\ref{su3rel4}) are satisfied to
within $15\,$MeV,
$13\,$MeV and $4\,$MeV, respectively.  The hyperfine ratio
(\ref{hqrel3}) is $(\Sigma_b^*-\Sigma_b)/(\Sigma_c^*-\Sigma_c)=0.38$,
and $\overline\Sigma_b-\Lambda_b$ is equal to
$\overline\Sigma_c-\Lambda_c$ to within $15\,$MeV.  This is better
agreement with the symmetries than we even have a right to expect.

Of course, this new proposal implies certain issues of its own.  The
most striking question is whether the new $\Sigma_c$ and $\Sigma_b$
states are already ruled out.  Consider the $\Sigma_c$, since much
more is known experimentally about charmed baryons.
The $\Sigma_c$ is an isotriplet, so it comes in the charge states
$\Sigma_c^0$,  $\Sigma_c^+$ and $\Sigma_c^{++}$.  With the proposed
mass, these states are too light to decay strongly, to
$\Lambda_c^++\pi$.  Instead, the $\Sigma_c^+$ will decay radiatively,
\be
  \Sigma_c^+\to\Lambda_c^++\gamma\,,\nonumber
\ee
while the others decay weakly, via channels such as
\bea
   \Sigma_c^{++,0}&\to&\Sigma^\pm+\pi^+\nonumber\\
   &\to&p+\pi^\pm+K_S\nonumber\\
   &\to&\Sigma^\pm+\ell^++\nu\,.\nonumber
\eea
The challenge, then, is either to find these states or conclusively to
rule them out.

We should also note that nonrelativistic constituent quark models
typically do not favor such light $\Sigma_c^{(*)}$ and $\Sigma_b{(*)}^*$ as I
have suggested here.  (See, for example, recent papers by
Lichtenburg~\cite{Lich} and Franklin.~\cite{Fran})  These models often have
been successful at predicting hadron masses, and are thus, not
unreasonably, quite popular.  However, despite common
misperceptions,~\cite{Lich,Fran} they are {\it less\/} general, and
make substantially {\it more\/} assumptions, than a treatment based
solely on heavy quark and $SU(3)$ symmetry.  A reasonable quark model
respects these symmetries in the appropriate limit, as well as
parametrizing deviations from the symmetry limit. Such models
therefore cannot be reconciled simultaneously with the heavy quark
limit and with the reported masses of the $\Sigma_b$ and
$\Sigma_b^*$.  Hence, the predictions of this analysis follow
experiment in pointing to physics beyond the constituent quark model.
While the historical usefulness of this model for hadron spectroscopy
may deepen one's suspicion of the DELPHI data on $\Sigma_{b1,2}$, such
speculation is beyond the scope of this discussion.  To reiterate, I
have taken the masses and errors of all states as they have been
reported to date; as they evolve in the future, so, of course, will
the theoretical analysis.

\section{Strong Decays of Excited Charmed Mesons}\label{STRONG}

Let us turn now from the spectroscopic implications of heavy quark
symmetry to its implications for the strong decays of excited
hadrons.  We will focus on the system for which there is the most, and
most interesting, data available, th excited charmed mesons.

As we saw in Section~\ref{SPEC}, there are two doublets of $p$-wave
charmed mesons, one with $J_\ell^P=\case12^+$ and one with
$J_\ell^P=\case32^+$.  The former correspond to the physical states
$D_0^*$ and $D_1'$, the latter to $D_1$ and $D_2^*$.  Note that the
$D_1$ and $D_1'$ both have $J^P=1^+$, being distinguished by their light
angular momentum $J_\ell^P$, which is a good quantum number only in
the limit $m_c\to\infty$.

The $D_0^*$ and $D_1'$ decay via $s$-wave pion emission,
\bea
  D^*_0&\to&D+\pi\nonumber\\
  D_1&\to&D^*+\pi\,.\nonumber
\eea
If their masses do not lie close to the threshold for this decay, then
these states can easily be quite broad, with widths of order 100 MeV
or more.  Hence they could be very difficult to identify
experimentally, and in fact no such states have yet been found.  By
contrast, the $D_1$ and $D_2^*$ are constrained by heavy quark symmetry to
decay via $d$-wave pion emission.  The channels which are allowed are
\bea\label{d12trans}
  D_1&\to&D^*+\pi\nonumber\\
  D_2^*&\to&D^*+\pi\nonumber\\
  D_2^*&\to&D+\pi\,.
\eea
Because their decays rates are suppressed by a power of $|{\bf
p}_\pi|^5$, these states could be much narrower than the $D_0^*$ and
$D_1'$.  In fact, resonances decaying in these channels have been
identified, and the properties of the $D_1(2420)$ and the $D_2(2460)$
are given in Table~\ref{mesontable}.

Since pion emission is a transition of the light degrees of freedom
rather than of the heavy quark, all of the decays (\ref{d12trans})
are really a single nonperturbative process, differentiated only by
the relative orientation of the spins of the heavy quark and the
initial and final light degrees of freedom.  Hence the three
transitions are related to each other by heavy quark symmetry.
In the strict limit $m_c\to\infty$, both the $D_1$ and $D_2^*$ and
the $D$ and $D^*$ are degenerate doublets, so the factor of $|{\bf
p}_\pi|^5$ is the same in all three decays.  The finite hyperfine
splittings $D^*-D\approx150\mev$ and $D_2^*-D_1\approx40\mev$ are
effects of order $1/m_c$, but their influence on $|{\bf p}_\pi|^5$ is
substantial.  Hence we will account for this factor explicitly by
invoking heavy quark symmetry at the level of the {\it matrix
elements\/} responsible for the decays, and using the physical
masses to compute the phase space.  A straightforward calculation
then yields two predictions for the full and partial widths:~\cite{ILW}
\bea
  &&{\Gamma(D_2^*\to D+\pi)/\Gamma(D_2^*\to D^*+\pi)}
  =2.3\label{d2pred}\\
  &&{\Gamma(D_1)/\Gamma(D_2^*)}=0.30\,.\label{d1pred}
\eea
For comparison, the experimental ratios are
\bea
  &&{\Gamma(D_2^*\to D+\pi)/\Gamma(D_2^*\to D^*+\pi)}
  =2.2\pm0.9\\
  &&{\Gamma(D_1)/\Gamma(D_2^*)}=0.71\,.
\eea
We see that the first relation works very well, while the second
fails miserably.  This unfortunate prediction raises a similar
question as we faced earlier:  is this a sign of a {\it general\/}
failure of heavy quark symmetry as applied to charmed mesons, or can
it be understood {\it within\/} the heavy quark expansion?  Naturally,
we would much prefer this latter outcome, for the familiar reason that
we want very much to believe we can trust this expansion for the
charmed mesons in other contexts.

Explanations for the failure for the prediction~(\ref{d1pred}) have
been offered in the past.  One is to suppose a small mixing,~\cite{ILW}
of order $1/m_c$, of the narrow $D_1$ with the broad $s$-wave $D_1'$.
Since these states have the same total angular momentum and parity,
$J^P=1^+$, such mixing is allowed when corrections for finite $m_c$ are
included.  A small mixing, of the size one might reasonably expect at
this order, could easily double the width of the physical $D_1$.  This
is a plausible explanation, and could well contribute at some level,
but for two reasons it is somewhat unlikely to be the dominant effect.
First, there is no evidence~\cite{CLEO} for an $s$-wave component in the
angular distribution of ${\bf p}_\pi$ in the decay $D_1\to
D^*+\pi$.  Although such a component could have escaped
undetected by a conspiracy of unknown final interaction phases, such a
situation is certainly not the generic one.  Second, there is no
evidence for an equivalent mixing between the strange analogues
$D_{s1}$ and $D_{s1}'$, which would broaden the observed $D_{s1}$
unacceptably.~\cite{ChoTriv}  Of course, light flavor $SU(3)$ might do
a poor job of predicting a mixing angle, which is actually a ratio of
matrix elements both of which receive $SU(3)$ corrections.  So, while
this explanation is not ruled out, neither does this evidence give one
particular confidence in it.

Another possibility is that the width of the $D_1$ receives a large
contribution from two pion decays to the $D$, either
nonresonant,~\cite{FalkLuke}
\be
  D_1\to D+\pi+\pi\,,\nonumber
\ee
or through an intermediate $\rho$ meson,~\cite{rhodecay}
\be
  D_1\to D+\rho\to D+\pi+\pi\,.\nonumber
\ee
Again, the problem is that there is no experimental evidence for such
an effect.  Also, it is somewhat difficult, within the schemes in which
such decays are discussed, to broaden the $D_1$ enough to match fully
the experimental width.  Hence, we are motivated to continue to search
for a more elegant and plausible explanation, which does not force us
to give up heavy quark symmetry for charmed mesons.

The answer, it turns out, lies in studying the heavy quark expansion
for the excited charmed mesons at subleading order in $1/m_c$. In this
case, we need a theory which contains both charmed mesons and soft
pions, coupled in the correct $SU(3)$ invariant way.  Such a
technology is heavy hadron chiral perturbation
theory.~\cite{HHCPT}  While the formalism is in some ways more than we
need, as it includes complicated pion self-couplings which will play
no role here, it is useful in that it allows us to keep track of all
the symmetries in the problem mechanically (and correctly).

Heavy hadron chiral perturbation theory accomplishes three things.
First, it builds in the heavy quark and chiral $SU(3)$ symmetries
explicitly.  Second, it implements a momentum expansion for the pion
field, in powers of $\partial_\mu\pi/\Lambda_\chi$, where the chiral
symmetry breaking scale is $\Lambda_\chi\approx 1\gev$.  Finally, and
very important in the present context, it allows one to include
symmetry breaking corrections in a {\it systematic\/} way.

To implement the symmetries, the Lagrangian must be built out of
objects which carry representations not just of the Lorentz group, but
of the heavy quark and $SU(3)$ symmetries as well.  Clearly, these
objects must contain both members of a single heavy meson doublet
of fixed $J_\ell^P$, and depend explicitly on the heavy meson
velocity.  For the ground state mesons $D$ and $D^*$, this the
``superfield''~\cite{FalkLuke,FGGW,traceform}
\be
  H_a = {(1+\rlap/v)\over2\sqrt2}
  \left[ D_a^{*\mu}\gamma_\mu-D_a\gamma^5\right]\,,
\ee
where the index on the $D^*$ is carried by the polarization vector.
Under heavy quark spin rotations $Q$, Lorentz transformations $L$, and
$SU(3)$ transformations $U$, $H_a$ transforms respectively as
\bea
  H_a&\to&S_QH_a\\
  H_a&\to&S_LH_aS_L^\dagger\\
  H_a&\to&H_aU_{ab}^\dagger\,.
\eea
Here $S_Q$ and $S_L$ are the spinor representations of the Lorentz
group, and $U_{ab}$ is the usual matrix representation of the vector subgroup
of spontaneously broken chiral $SU(3)$ symmetry. There are similar
superfields for the excited mesons,~\cite{FalkLuke,traceform}
\bea
  S_a &=& {(1+\rlap/v)\over2\sqrt2}
  \left[ D_{1a}^{\prime\mu}\gamma_\mu\gamma^5-D_{0a}^*\right]\,,\\
  T^\mu_a &=& {(1+\rlap/v)\over2\sqrt2}\left[
  D_{2a}^{*\mu\nu}\gamma_\nu
  -D_{1a}^\nu\sqrt{\case32}\,\gamma^5(\delta^\mu_\nu-\case13
  \gamma_\nu\gamma^\mu+\textstyle{1\over3}\gamma_\nu v^\mu)\right]\,,
\eea
transforming in analogous ways.  The superfields $H_a$, $S_a$ and $T^\mu_a$ all
have mass dimension $\case32$.  The pion fields appear as in
ordinary chiral perturbation theory; since we will not be interested in
pion self-couplings, we will just recall the linear term in the
exponentiation of the pion fields,
\be
  A_\mu={1\over f_\pi}\partial_\mu{\Pi}+\dots\,,
\ee
where ${\Pi}$ is the matrix of Goldstone boson fields and
$f_\pi\approx132\mev$.

We now build a Lagrangian out of invariant combinations of these
elements.  At leading order we get the terms responsible for the $d$-wave
decay of the $D_1$ and $D_2^*$,
\be
  {h\over\Lambda_\chi}\,{\rm Tr}\,\left[\overline H\,
  T^\mu\gamma^\nu\gamma^5
  (iD_\mu A_\nu+iD_\nu A_\mu)\right]+{\rm h.c.}\,,
\ee
and for the
$s$-wave decay of the $D_0^*$ and $D_1'$,
\be
  f\,{\rm Tr}\,\left[\overline H\,
  S\gamma^\mu\gamma^5A_\mu\right]
  +{\rm h.c.}\,.
\ee
Expanding these interactions in terms of the individual fields, we find
the same symmetry predictions (\ref{d2pred}) and (\ref{d1pred}) as
before.

However, now we would like to go further, and include the leading
corrections of order $1/m_c$ in the effective
Lagrangian.~\cite{FaMe95}  To understand how to do this, we turn to the
expansion of QCD in the heavy quark limit, given by the heavy quark
effective theory.  This Lagrangian is written in terms of an effective
HQET field $h(x)$, which satisfies the conditions~\cite{Georgi}
\be
  {1+\rlap/v\over2}h(x)=h(x)
\ee
and
\be
  i\partial_\mu h(x)=k_\mu h(x)\,,
\ee
where $k^\mu=p_c^\mu-m_cv^\mu$ is the ``residual momentum'' of the
charm quark.  Including the leading corrections, the HQET Lagrangian
takes the form~\cite{Georgi,FGL}
\be
   {\cal L}_{\rm HQET} = \bar hiv\cdot Dh+{1\over2m_c}\bar h(iD)^2h
   +{1\over2m_c}\bar h\sigma^{\mu\nu}(\textstyle{1\over2}gG_{\mu\nu})h
   +\dots\,.
\ee
The effect of the subleading terms $\bar h(iD)^2h$ and $g\bar
h\sigma^{\mu\nu}G_{\mu\nu}h$ on the chiral expansion may be treated in
the same manner as other symmetry breaking perturbations to the
fundamental theory such as finite light quark masses.  Namely, we
introduce a ``spurion'' field which carries the same representation of
the symmetry group as does the perturbation in the fundamental theory,
and then include this spurion in the chiral lagrangian in the most
general symmetry-conserving way.  When the spurion is set to the
constant value which it has in QCD, the symmetry breaking is
transmitted to the effective theory.  In the case of finite light
quark masses, for example, the symmetry breaking term in QCD is $\bar
q M_qq$, where $M_q={\rm diag}(m_u,m_D,m_s)$.  Introducing a spurion
$M_q$ which transforms as $M_q\to LM_qR^\dagger$ under chiral $SU(3)$,
we then include terms in the ordinary chiral lagrangian such as
$\mu\,{\rm Tr}\,[M_q\Sigma+M_q\Sigma^\dagger]$.

In the present case, only the second of the two correction terms in
${\cal L}_{\rm HQET}$ violates the heavy spin symmetry.  We include
its effect in the chiral lagrangian by introducing a spurion
$\Phi_s^{\mu\nu}$ which transforms as $\Phi_s^{\mu\nu}\to
S_Q\Phi_s^{\mu\nu}S_Q^\dagger$ under a heavy quark spin rotation
$S_Q$.  This spurion is introduced in the most general manner
consistent with heavy quark symmetry, and is then set to the constant
$\Phi_s^{\mu\nu}=\sigma^{\mu\nu}/2m_c$ to yield the leading spin
symmetry violating corrections to the chiral lagrangian.  We will
restrict ourselves to terms in which $\Phi_s^{\mu\nu}$ appears exactly
once.

The simplest spin symmetry violating effect is to break the degeneracy
of the heavy meson doublets.  This occurs through the terms
\be
   \lambda_H\,{\rm Tr}\,\left[\overline H\Phi_s^{\mu\nu}H\sigma_{\mu\nu}
   \right]
   -\lambda_S\,{\rm Tr}\,\left[\overline S\Phi_s^{\mu\nu}S
   \sigma_{\mu\nu}\right]
   -\lambda_T\,{\rm Tr}\,\left[\overline T^\alpha\Phi_s^{\mu\nu}T_\alpha
   \sigma_{\mu\nu}\right]\,.
\ee
The dimensionful coefficients are fixed once the masses of the mesons
are known.  For the ground state $D$ and $D^*$, for example, we find
\be
   \lambda_H={1\over8}\left[M^2_{D^*}-M^2_D\right]=(260\,{\rm MeV})^2\,.
\ee
This value is entirely consistent with what one would obtain, instead,
with the $B$ and $B^*$ mesons.  For the $D_1$ and $D_2^*$, we find
\be
   \lambda_T={3\over16}\left[M^2_{D_2^*}-M^2_{D_1}\right]=(190\,{\rm MeV})^2\,.
\ee
Note that $\sqrt{\lambda_H}$ and $\sqrt{\lambda_T}$ are of order
hundreds of MeV, the scale of the strong interactions.

We are interested in the spin symmetry violating corrections to
transitions in the class $T^\mu\to H+\pi$, which will arise from terms
analogous to ${\cal L}_d$ but with one occurrence of
$\Phi_s^{\mu\nu}$.  The spin symmetry, along with the symmetries which
constrained ${\cal L}_d$, requires that any such term be of the
generic form
\be
   {1\over\Lambda_\chi}{\rm Tr}\,\left[\overline H\Phi_s^{\mu\nu}T^\alpha
   C_{\mu\nu\alpha\beta\kappa}\gamma^5
   \left(iD^\beta A^\kappa+iD^\kappa A^\beta\right)\right]+{\rm h.c.}\,,
\ee
where $C_{\mu\nu\alpha\beta\kappa}$ is an arbitrary product of Dirac
matrices and may depend on the four-velocity $v^\lambda$.  This would
seem to allow for a lot of freedom, but it turns out that there is
only a {\it single\/} spin symmetry-violating term which respects both parity
and time
reversal invariance:
\be
   {\cal L}_{d1} = {h_1\over2m\Lambda_\chi}{\rm Tr}\,
   \left[\overline H\sigma^{\mu\nu}T^\alpha
   \sigma_{\mu\nu}\gamma^\kappa\gamma^5
   \left(iD_\alpha A_\kappa+iD_\kappa A_\alpha\right)\right]+{\rm h.c.}\,.
\ee
We expect the new coefficient $h_1$, which has mass dimension one, to
be of order hundreds of MeV.

The mixing of $D_1$ and $D_1'$ is also a spin symmetry violating
effect which arises at order $1/m_c$.  There is a corresponding
operator in the chiral lagrangian which is responsible for this,
\be\label{Lmix}
   {\cal L}_{\rm mix} = g_1{\rm Tr}\,\left[\overline S\Phi_s^{\mu\nu}T_\mu
   \sigma_{\nu\alpha}v^\alpha\right]+{\rm h.c.}\,.
\ee
However, we will neglect this term for now.  It is straightforward to
include both ${\cal L}_{d1}$ and ${\cal L}_{\rm mix}$ in a more
complete analysis.~\cite{FaMe95}

We now compute the partial widths for the decays of the $D_1$ and the
$D_2^*$ at subleading order in the $1/m_c$ expansion.  We find
\bea\label{d2widths}
   \Gamma(D_2^{*0}\to D\pi)&=&{1\over10\pi}\,{m_D\over M_{D_2^*}}
   \,{4|{\bf p}_\pi|^5\over
   \Lambda_\chi^2f_\pi^2}\left[h-{h_1\over m_c}\right]^2\\
   \Gamma(D_2^{0*}\to D^*\pi)&=&{3\over20\pi}\,{M_{D^*}\over M_{D_2^*}}
   \,{4|{\bf p}_\pi|^5\over
   \Lambda_\chi^2f_\pi^2}\left[h-{h_1\over m_c}\right]^2\\
   \Gamma(D_1\to D^*\pi)&=&{1\over4\pi}\,{M_{D^*}\over M_{D_1}}
   \,{4|{\bf p}_\pi|^5\over
   \Lambda_\chi^2f_\pi^2}\left[\left(h+{5h_1\over 3m_c}\right)^2
   +{8h_1^2\over9m_c^2}\right]\,,
\eea
where in each expression $|{\bf p}_\pi|^5$ is computed using the
actual phase space for that decay.  Setting $h_1=0$ would reduce these
results to the leading order predictions.  Note that the ratio of
partial widths of the $D_2^*$ is independent of $h_1$, and so is {\it
unchanged\/} by the inclusion of $1/m_c$ effects.  However, the ratio
of the widths of the $D_1$ and the $D_2^*$  receives a large
correction,
\be
  {\Gamma(D_1)/\Gamma(D_2^*)}=0.30\left[1+{16\over3}{h_1\over m_c h}
  +\dots\right]\,.
\ee
{}From the width of the $D_2^*$, and taking $\Lambda_\chi=1\gev$, we
find $h\approx 0.3$.  Then we see that even for a modest coefficient
$h_1\approx100\mev$, we get a correction to the ratio of widths of
order 100\%!

What we have learned, then, is that a $1/m_c$ correction of the
canonical size, with no tuning of parameters, naturally leaves one of
these predictions alone while destroying the other.  In this sense, we
understand the failure of the bad prediction {\it within\/} the heavy
quark expansion.  This is what we mean by saying that heavy quark
symmetry (or any symmetry) ``works''.  It need not be the case that
every prediction of the symmetry limit be  well satisfied by the
data.  Rather, it is crucial that deviations from the symmetry limit
can be understood {\it within a systematic expansion in the small
parameters which break the symmetry.}  When a symmetry works in this
sense, we retain predictive power even in cases when the symmetry predictions
behave poorly.

\section{Production of Heavy Hadrons via Fragmentation}\label{FRAG}

Before heavy hadrons can decay, they must be produced.  The
production of a heavy hadron proceeds in two steps.   First, the
heavy quark itself must be created; because of its large mass, this
process takes place over a time scale which is very short.  Second,
some light degrees of freedom assemble themselves about the heavy
quark to make a color neutral heavy hadron, a process which involves
nonperturbative strong interactions and typically takes much
longer.  If the heavy quark is produced with a large velocity in the
center of mass frame, and if there is plenty of available energy,
then production of these light degrees of freedom will be local in
phase space and independent of the light degrees of freedom in the
initial state.  This is the fragmentation regime.  We will see that
heavy quark symmetry simplifies the description of heavy hadron
production via fragmentation, because, as before, it allows us to
separate certain properties of the heavy quark from those of
the light degrees of freedom.  This is particularly important in the
production of excited heavy hadrons, for which the behavior of the
spin of the light degrees of freedom can be quite interesting.

Our consideration of heavy quark fragmentation will lead us to
consider two related questions:~\cite{FaPe94}
\par\noindent 1.~What are the nonperturbative features of the
fragmentation process?  In particular, can we exploit heavy quark
symmetry to isolate and study the spin of the light degrees of
freedom?
\par\noindent 2.~What is the fate of a polarized heavy quark created
in the hard interaction?  Is any initial polarization preserved
until the heavy quark undergoes weak decay?
\par\noindent We will see that an understanding of the first
question will cast a useful light on the second.  In the latter
case, the excited heavy baryons will play a significant role.

The analysis depends on following the spins of the heavy quark and
the light degrees of freedom separately through the three phases of
fragmentation, life of the state, and decay.  The net interaction of
the heavy and light angular momenta $S_Q$ and $J_\ell$ depends both
on the strength of the coupling between them and on the length of
time they have to interact.  Of course, the coupling between the
spins is small in the heavy quark limit, because it is mediated by
the chromomagnetic moment of the heavy quark.  This moment scales as
$1/m_Q$, so the time $\tau_s$ it take for the heavy and light spins
to precess once about each other is of order $m_Q/\lqcd^2$, much
longer than typical time scales associated with the strong
interactions.

This fact is enough to assure that the heavy quark spin is
essentially frozen during the process of fragmentation itself.
Since fragmentation is purely a phenomenon of nonperturbative QCD,
it takes place on a time scale of order $1/\lqcd\ll\tau_s$.  Hence
there is not enough time for the relatively weak spin-exchange
interactions to take place.

Naively, one can say something similar when the heavy quark fragments
to an excited hadron which decays via a strong transition of the
light degrees of freedom.  The time scale of a strong transition is
set by nonperturbative QCD and should be comparable to the
fragmentation time.  Thus, one might expect generically that the
lifetime $\tau$ of the state satisfies $\tau\ll\tau_s$, and the
heavy quark spin continues to be frozen in place during the life of
the excited hadron.  However, if the energy available in the decay
is not much larger than $m_\pi$, the lightest hadron which can be
emitted in a strong transition, then $\tau$ can be increased by the
limited phase space.  The most dramatic example is $D^*$ decay,
which is so close to threshold that the strong ($D^*\to D+\pi$) and
electromagnetic ($D^*\to D+\gamma$) widths are almost equal.

So we must treat excited hadrons on a case by case basis, depending
on the relative sizes of $\tau$ and $\tau_s$.  For simplicity, we
will consider here only two extreme cases.  Let the excited heavy
doublet be composed of a hadron $H$ of spin $J$ and mass $M$ and a
hadron $H^*$ of spin $J+1$ and mass $M^*$.   The first possibility is
the ``naive'' one $\tau_s\gg\tau$, where $H$ and $H^*$ are formed
and then decay before the angular momenta $S_Q$ and $J_\ell$ have a chance to
interact.  In this case, there is no depolarization of the heavy
quark spin $S_Q$, if one was present initially.  Similarly, when
$H$ and $H^*$ decay strongly, the light degrees of freedom in the
decay carry any information about the spin state in which they
were produced.  Note that the very spin-exchanges interaction which
is inhibited here is the one responsible for the hyperfine
splitting between $H$ and $H^*$.  Hence, under these conditions the resonances
are almost completely {\it overlapping,} with a width
$\Gamma=1/\tau$ satisfying $\Gamma\gg|M^*-M|$.  This is another
consequence of the effective decoupling of $S_Q$ and $J_\ell$,
which are independent good quantum numbers of the resonances.

The second possibility is the opposite extreme, $\tau\gg\tau_s$.
This corresponds to heavy hadrons which decay weakly or
electromagnetically, or to strong decays which are severely
suppressed by phase space.  Here the spins $S_Q$ and $J_\ell$ have
plenty of time to interact, precessing about each other many times
before $H$ and $H^*$ decay.  There is at least a partial degradation
of any initial polarization of $Q$, as well as a degradation of any
information about the fragmentation process which may be carried by
the light degrees of freedom.  The signature of this situation is
that the states $H$ and $H^*$ are well separated resonances, since
the chromomagnetic interactions have ample opportunity to produce a
hyperfine splitting much larger than the width,
$|M^*-M|\gg\Gamma$.  In contrast with the first case, here the
heavy and light spins are resolved into states of definite total
spin $J$.

\subsection{Production and decay of $D_1$ and $D_2^*$}

We will consider two examples, the first of which is the production
and decay of the excited charmed mesons $D_1$ and $D_2^*$.  We see
from Table~\ref{mesontable} that the splitting between these
states is 35~MeV, while their widths are approximately 20~MeV.
This makes them somewhat of an intermediate case; however, for
simplicity let us treat them in the ``widely separated resonances''
limit.  A more precise treatment which takes into
account their finite widths is straightforward but not very
pedagogically enlightening.~\cite{FaMe95,FaPe94}

We must follow the orientations of the spins $S_Q$ and $J_\ell$
through the following sequence of events:
\par\noindent 1.~The charm quark is created in some hard
interaction.
\par\noindent 2.~Light degrees of freedom with $J_\ell^P=\case32^+$
are created in the process of fragmentation.
\par\noindent 3.~The spins $S_Q$ and $J_\ell$ precess about each
other, resolving the states $D_1$ and $D_2^*$ of definite total
angular momentum $J$.
\par\noindent 4.~The $D_1$ or the $D_2^*$ decays via $d$-wave pion
emission.  We can measure the direction of this pion with respect
to the spatial axis along which the fragmentation took place.
\par\noindent The light degrees of freedom can be produced with
helicity $h=\pm\case32$ or $h=\pm\case12$ along the fragmentation
axis.  While parity invariance of the strong interactions requires
that the probabilities for helicities $h$ and $-h$ are identical,
the relative production of light degrees of freedom with
$|h|=\case32$ versus $|h|=\case12$ is determined in some
complicated and incalculable way by strong dynamics.  Let the
quantity $w_{3/2}$ denote the probability that $|h|=\case32$,
\be
  w_{3/2}=P(h=\case32)+P(h=-\case32)\,.
\ee
Then $1-w_{3/2}$ is the probability that $|h|=\case12$.  Completely
isotropic production corresponds to $w_{3/2}=\case12$.  We have identified a
new nonperturbative parameter of QCD, which is well defined only in
the heavy quark limit.

This new parameter can be measured in the strong decay of the
$D_2^*$ or $D_1$.  For example, consider the angular distribution
of the pion with respect to the fragmentation axis in the decay
$D_2^*\to D+\pi$.  This is a decay of the light degrees of freedom
in the excited hadron, so it will depend on their initial
orientation (that is, on $w_{3/2}$) and on the details of the
precession of $J_\ell$ around $S_Q$ during the lifetime of the
$D_2^*$.  Following the direction of $J_\ell$ through
fragmentation, precession and decay, we find the distribution
\be\label{d2todpi}
  {1\over\Gamma}{{\rm d}\Gamma\over{\rm d}\cos\theta}=
  \case14\left[1+3\cos^2\theta-6w_{3/2}
  (\cos^2\theta-\case13)\right]\,.
\ee
This distribution is isotropic only when $w_{3/2}=\case12$, that
is, when the light degrees of freedom are produced isotropically in
the fragmentation process.  Similar distributions are found in the
decays $D_2^*\to D^*+\pi$ and $D_1\to D^*+\pi$.

A fit of ARGUS data~\cite{ARGUS} to the expression (\ref{d2todpi}) seems
to indicate that a small value of $w_{3/2}$ is preferred; while the
errors are large, we find that
$w_{3/2}<0.24$ at the 90\% confidence level.~\cite{FaPe94}  It would be
nice to confirm this result with a sharper measurement, and not only for
the charmed mesons but in the bottom system as well.  Since $w_{3/2}$
is intrinsically nonperturbative, we do not have any real
theoretical understanding of why it should be small, although
perturbative calculations of of fragmentation production of the $B_c$
system in the limit $m_c\ll m_b$ yield small $w_{3/2}$ as
well.~\cite{ChenWise,Yuan}

\subsection{Polarization of $\Lambda_b$ at SLC/LEP}

After warming up with the excited charmed mesons, we are set to
address a somewhat more practical question:  What is the polarization of
$\Lambda_b$ baryons produced at the $Z$ pole?  This question is
motivated by the fact that $b$ quarks produced in the decay of the
$Z$ are 94\% polarized left-handed.  Since the $\Lambda_b$ is
composed of a $b$ quark and light degrees of freedom with zero
net angular momentum, the orientation of a $\Lambda_b$ is identical
to the orientation of the $b$ quark inside it.  Similarly, the $b$
quark spin does not precess inside a $\Lambda_b$.  Hence if a
$b$ quark produced at the $Z$ fragments to a $\Lambda_b$, then
those baryons should inherit the left-handed polarization of the
quarks and reveal it in their weak decay.

Unfortunately, life is not that simple.  Two recent measurements of
$\Lambda_b$ polarization from LEP are~\cite{DELPHI1,ALEPH}
\bea
  &&P(\Lambda_b) =0.08^{+0.35}_{-0.29}{\rm (stat.)}^{+0.18}_{-0.16}
  {\rm (syst.)}\qquad {\rm (DELPHI)}\,,\nonumber\\
  &&P(\Lambda_b) =0.26^{+0.20}_{-0.25}{\rm (stat.)}^{+0.12}_{-0.13}
  {\rm (syst.)}\qquad {\rm (ALEPH)}\,,\nonumber
\eea
both a long way from $P(\Lambda_b) =0.94$.  The reason is that not
all $b$ quarks which wind up as $\Lambda_b$ baryons get there
directly.  In particular, they can fragment to the excited baryons
$\Sigma_b$ and $\Sigma_b^*$, which then decay to
$\Lambda_b$ via pion emission.  If the excited states, which
have light degrees of freedom with $S_\ell=1$, live long enough,
then the $b$ quark will precess about $S_\ell$ and the polarization
will be degraded.  The result will be a net sample of $\Lambda_b$'s
with a polarization less than 94\%, as is in fact observed.

In addition to the requirement that $\tau>\tau_s$ for the
$\Sigma_b^{(*)}$, any depolarization of $\Lambda_b$'s by this
mechanism depends on two unknown quantities:
\par\noindent 1.~The production rate $f$ of $\Sigma_b^{(*)}$ relative
to $\Lambda_b$.  Isospin and spin counting enhance $f$ by a factor
of nine, while the mass splitting between $\Sigma_b^{(*)}$ and
$\Lambda_b$ suppresses it; studies based on the Lund Monte Carlo
indicate $f\approx0.5$ with a very large uncertainty.~\cite{LUND}
\par\noindent 2.~The orientation of the spin $S_\ell$ with respect
to the fragmentation axis.  This orientation, which is
nonperturbative in origin, reflects the possible helicities
$h=1,0,-1$.  By analogy with the treatment of the heavy mesons, we
define~\cite{FaPe94}
\be
  w_1=P(h=1)+P(h=-1)\,.
\ee
In this case, isotropic production corresponds to $w_1=\case23$.  We
may measure $w_1$ from the angle of the pion with respect to the fragmentation
axis in the decay
$\Sigma_b^*\to\Lambda_b+\pi$,
\be\label{sigtolampi}
  {1\over\Gamma}{{\rm d}\Gamma\over{\rm d}\cos\theta}=
  \case14\left[1+3\cos^2\theta-\case92w_1
  (\cos^2\theta-\case13)\right]\,.
\ee
It turns out that the decay $\Sigma_b\to\Lambda_b+\pi$ is isotropic in
$\cos\theta$ for any value of $w_1$.

The polarization retention of the $\Lambda_b$ may be computed in
terms of $f$ and $w_1$.  As before, it is more tedious than
instructive to present the general case in which the $\Sigma_b$ and
the $\Sigma_b^*$ may partially overlap, so let us restrict to the
extreme situation $\tau\gg\tau_s$.  Then the polarization of the
observed $\Lambda_b$'s is $P(\Lambda_b)=R(f,w_1)P(b)$, where
$P(b)=94\%$ is the initial polarization of the $b$ quarks,
and~\cite{FaPe94}
\be
  R(f,w_1) = {1+\case19(1+4w_1)f\over1+f}\,.
\ee
Note that for $f=0$ (no $\Sigma_b^{(*)}$'s are produced),
$R(0,w_1)=1$ and there is no depolarization.  For the Lund value
$f=0.5$, $R$ ranges between 0.70 and 0.85.

Can the very low measured values of $P(\Lambda_b)$ be accommodated
by the present data on the $\Sigma_b^{(*)}$?  The situation is
still unclear.  On the one hand, the same DELPHI analysis which found
such surprising masses for the excited bottom baryons reported
$w_1\approx0$ and $1<f<2$ with large uncertainty.~\cite{DELPHI}  If this
is confirmed, and if the conventional identification of the bottom
baryons is correct, then a polarization in the range
$P(\Lambda_b)\approx40\%-50\%$ is easy to accommodate.  On the other
hand, CLEO's recent announcement~\cite{CLEO96} of the $\Sigma_c^*$ was
accompanied by a measurement $w_1=0.71\pm0.13$, consistent with
isotropic fragmentation.  Recall that by heavy quark symmetry, $w_1$
measured in the charm and bottom systems must be the same, so this
result is inconsistent with the report from DELPHI.  Clearly, further
measurements are needed to resolve this situation.

\section{Weak Decays}\label{WEAK}

We now turn to our final topic, the production of excited charmed
hadrons in semileptonic $B$ decays.  The branching fraction of
\be
  B\to(D_1,D_2^*)+\ell+\nu
\ee
has been measured by two groups with roughly
consistent results:~\cite{branch}
\bea\label{D12data}
  {\rm OPAL}:&&\qquad(34\pm7)\%\\
  {\rm CLEO}:&&\qquad<30\%\ {\rm at\ 90\%\ c.l.}\,.
\eea
It is not unreasonable to assume that this measurement will
eventually be improved, and any discrepancies resolved.  The
question is, what can we learn from it?  How useful would an effort
to improve this measurement really be?

I will propose that it would be extremely useful.  First, because
studying the production of  excited charmed mesons in $B$ decay
gives us direct information about QCD, and second, because through
this insight into QCD we can dramatically reduce the single most
nettlesome theoretical uncertainty in the extraction of $\vcb$ from
inclusive semileptonic $B$ decays, namely the dependence on the $b$ quark mass.

The heavy quark expansion and perturbative QCD may be used to analyze
semileptonic and radiative $B$ decays in a systematic expansion in
powers of $1/m_b$ and $\alpha_s(m_b)$.~\cite{CGG,SV,FLS94,MW}  Since the
energy $m_b-m_c$ which is released in such a decay is large compared to
$\lqcd$, we may invoke the duality of the partonic and hadronic
descriptions of the process.  The idea is that sufficiently inclusive
quantities may be computed at the level of quarks and gluons, if the
interference between the short-distance and long-distance physics may
be neglected.  Except near the boundaries of phase space, this is
usually the case if the ratio of typical long wavelengths
($\sim1/\lqcd$) to typical short wavelengths ($\sim1/m_b$) is
sufficiently large.  While it is reasonable to expect parton-hadron
duality to hold for arbitrarily large  energy releases, its
application at the $b$ scale requires a certain amount of
faith.~\cite{FDW}

Consider a $B$ meson with initial momentum $p_B^\mu=m_B v^\mu$, which
decays into leptons with total momentum $q^\mu$ and a hadronic state
$X_c$ with momentum $p_X^\mu=p_B^\mu-q^\mu$.  Since we are interested
in the properties of the hadrons which are produced, we define the
kinematic invariants~\cite{FLS96}
\bea
  s_H&=&p_X^2\\
  E_H&=&p_X\cdot p_B/m_B\,,
\eea
which are, respectively, the invariant mass of the hadrons and their
total energy in the $B$ rest frame.  We then compute the doubly
differential distribution $\d\Gamma/\d s_H\d E_H$ using the heavy
quark expansion.  First, we use the optical theorem to relate the
semileptonic decay rate for fixed $q^\mu$ to the imaginary part of a
forward scattering amplitude,
\bea
  &&\sum_{X_c}\int\d q\,\big|\langle X_c(p_X)(\ell\nu)(q)|
  {\cal O}_W|B\rangle\big|^2\nonumber\\
  &&\qquad
  =\case12 G_F^2\int\d q\,e^{iq\cdot x} L_{\mu\nu}(q)\,\langle B|
  T\{J_{bc}^{\dagger\mu}(x),J_{bc}^\nu(0)\}|B\rangle\,.
\eea
Here ${\cal O}_W=(G_F/\sqrt2)J_{cb}^\mu J_{\ell\mu}$ is the product of
left-handed currents responsible for the semileptonic decay $b\to
c\ell\nu$, and
\be
  L_{\mu\nu}=\case13\left(q_\mu q_\nu-q^2g_{\mu\nu}\right)
\ee
is the tensor derived from the squared leptonic matrix element. The next step
is to  expand the time-ordered product
$T\{J_{bc}^{\dagger\mu}(x),J_{bc}^\nu(0)\}$ in inverse powers of
$1/m_b$, using the operator product expansion and the heavy quark
effective theory.  This yields an infinite sum of operators written in
terms of the effective field $h(x)$, which we will truncate at order
$1/m_b^2$.  Finally, we write the matrix elements of the form
$\langle B|\bar h\cdots h|B\rangle$ in terms of parameters given by the heavy
quark expansion.

Once we have the differential distribution $\d\Gamma/\d s_H\d E_H$, we
can weight with powers of the form $s_H^n E_H^m$ and integrate to
compute moments of $s_H$ and $E_H$.  Of course the $(n,m)=(0,0)$
moment is just the semileptonic partial width $\Gamma$.  The moments
of $s_H$, which will be of particular interest, are sensitive to the
production of excited charmed hadrons such as the $D_1$ and $D_2^*$.

Our results will be in terms of four QCD and HQET parameters, since we
keep only terms up to order $1/m_b^2$:
\par\noindent 1.~The strong coupling constant $\alpha_s(m_b)$.  We get
powers of $\alpha_s(m_b)/\pi$ when we compute the radiative corrections
to the time-ordered product.
\par\noindent 2.~The ``mass'' $\bar\Lambda$ of the light degrees of
freedom, defined by~\cite{Luke}
\be
  \bar\Lambda=\lim_{m_b\to\infty}\left[m_B-m_b\right]\,.
\ee
Because the quark mass which appears in this expression is the pole
mass, $m_b=m_b^{\rm pole}$, the quantity $\bar\Lambda$ suffers from an
infrared renormalon ambiguity~\cite{renormalons} of order  $\sim100\mev$.
This ambiguity affects the interpretation of $\bar\Lambda$, and so we
must treat with caution any expression in which it appears.  For
comparison with data, it is preferable to use expressions in which the
renormalon ambiguity can be shown to cancel.
\par\noindent 3.~The ``kinetic energy'' $\lambda_1$ of the heavy quark,
defined by~\cite{FaNe}
\be
  \lambda_1=\lim_{m_b\to\infty}\langle B|\bar b(iD)^2 b|B\rangle/2m_B\,.
\ee
Note that $\lambda_1$ is not exactly the $b$ quark kinetic energy (or
rather, its negative), since there are gauge fields in the covariant
derivative.  Relative to the $b$ quark's rest energy, its
nonrelativistic kinetic energy is suppressed by $1/m_b^2$.
\par\noindent 4.~The energy of the $b$ quark due to its hyperfine
interaction with the light degrees of freedom, given by~\cite{FaNe}
\be
  \lambda_2=\lim_{m_b\to\infty}\langle B|\case12 g
  \bar b\sigma^{\mu\nu}G_{\mu\nu}b|B\rangle/6m_B\,.
\ee
This is the only one the four parameters where the spin of the $b$
enters directly.  We can extract $\lambda_2$ from the $B^*-B$ mass
splitting, which yields\,\footnote{Because the chromomagnetic operator
is renormalized, $\lambda_2(\mu)$ actually depends slightly on the
renormalization scale.~\cite{FGL,EiHill}  The number we give here is
$\lambda_2(m_b)$.}
\be
  \lambda_2=0.12\gev\,.
\ee

We will present results which include heavy quark corrections up
through order $1/m_b^2$, and radiative corrections up through two
loops.  Actually, the two loop corrections are only partially
computed, with just those pieces proportional to $\beta_0\alpha_s^2$,
where $\beta_0=11-\case23n_f$ is the leading coefficient in the QCD
beta function.  We may hope that this piece dominates the two loop
term, because of the large numerical coefficient $\beta_0$; in fact,
for other calculations for which the full two loop result is known,
this is usually the case.  For semileptonic $B$ decay, the
full two loop calculation has not been completed.

We will present results for the semileptonic partial width, and for
the first moment of the hadronic invariant mass spectrum.  It is
convenient to substitute all appearances of the charm and bottom quark
masses with spin-averaged meson masses, using the expansion
\be
  \overline m_B=m_B+\bar\Lambda-{\lambda_1\over2m_B}+\dots\,,
\ee
and analogously for charm.  Then the coefficients which appear below
are functions of the measured ratio $\overline m_D/\overline m_B$,
with no hidden dependence on unknown quark masses.  For the
semileptonic partial width, we find~\cite{FLS96}
\bea\label{width}
  \Gamma(B\to X_c\ell\nu)&=&{G_F^2\vcb^2\over192\pi^3}m_B^5\, 0.369
  \bigg[1-1.54{\alpha_s(m_b)\over\pi}-1.43\beta_0
  {\alpha_s^2(m_b)\over\pi^2}\nonumber\\
  &&\qquad\qquad-1.65{\bar\Lambda\over m_B}\left(1-0.87{\alpha_s(m_b)
  \over\pi}\right)-0.95{\bar\Lambda^2\over m_B^2}\nonumber\\
  &&\qquad\qquad-3.18{\lambda_1\over m_B^2}+0.02{\lambda_2\over m_B^2}
  +\dots\bigg]\,,
\eea
and for the average hadronic invariant mass,~\cite{FLS96}
\bea\label{moment}
  \langle s_H-\overline m_D^2\rangle &=& m_B^2\bigg[
  0.051{\alpha_s(m_b)\over\pi}+0.096\beta_0{\alpha_s^2(m_b)\over\pi^2}
  \nonumber\\
  &&\qquad\qquad+0.23{\bar\Lambda\over m_B}\left(1+0.43{\alpha_s(m_b)
  \over\pi}\right)+0.26{\bar\Lambda^2\over m_B^2}\nonumber\\
  &&\qquad\qquad+1.01{\lambda_1\over m_B^2}-0.31{\lambda_2\over m_B^2}
  +\dots\bigg]\,.
\eea
We include a subtraction of $\overline m_D^2$ in the invariant mass so
that the theoretical expression will start at order $\alpha_s$ and
$\bar\Lambda$. The heavy quark expansion seems to be under control, as
the corrections proportional to  $\lambda_1$ and $\lambda_2$ are at
the level of a few percent.  However, this not true of the expansion
in perturbative QCD.  Since $\beta_0\alpha_s/\pi\approx0.6$, we see
that the two loop corrections to (\ref{width}) and (\ref{moment}) are
as large as the one loop terms.

This is real trouble!  With such a poorly behaved perturbation series,
these expressions are not trustworthy.  Actually, there is a problem
with the nonperturbative corrections, too, since they contain the
ambiguous parameter $\bar\Lambda$.  How, then, can we use this theory
to do reliable phenomenology?

Remarkably, these two problems are actually connected, and can be used
to solve each other.  The renormalon ambiguity of $\bar\Lambda$ arises
from the poor behavior of QCD perturbation theory at high orders in
the series for $m_b^{\rm pole}$.  Perhaps it is the same poor
behavior which manifests itself in the perturbation series for
$\Gamma$ and $\langle s_H\rangle$.  If so, then the solution is to
eliminate $\bar\Lambda$ in favor of some unambiguous {\it physical\/}
quantity, solving both problems as once.

In fact, it can be shown that this is precisely the
case.~\cite{renormalons}  The bad perturbation series in $\Gamma$ arises
from the indirect dependence of the theoretical expression on the pole
mass $m_b^{\rm pole}$, through $\bar\Lambda$.  One way to eliminate
$\bar\Lambda$ is to write it in terms of $\langle s_H-\overline
m_D^2\rangle$, which can be measured.  We then find~\cite{FLS96}
\bea\label{width2}
  \Gamma(B\to X_c\ell\nu)&=&{G_F^2\vcb^2\over192\pi^3} m_B^5\, 0.369
  \bigg[1-1.17{\alpha_s(m_b)\over\pi}-0.74\beta_0
  {\alpha_s^2(m_b)\over\pi^2}\nonumber\\
  &&\qquad\qquad\qquad\qquad\qquad-7.17{\langle s_H-\overline m_D^2\rangle
  \over m_B^2}+\dots\bigg]\,,
\eea
omitting the small terms of order $1/m_b^2$.  Note that the size of
the two loop term has shrunk by a factor of two with this
rearrangement.  We have regained some measure of control over the
perturbation series.\footnote{This improvement may be interpreted as
an increase in the BLM renormalization scale~\cite{BLM} from $\mu_{\rm
BLM}=0.16m_B$ to $\mu_{\rm BLM}=0.38m_B$.~\cite{FLS96}}

The moral of this exercise is that while it is perfectly fine to keep
$\bar\Lambda$ in intermediate steps in calculations, it should be
eliminated from predictions of physical quantities.  By the same
token, any extraction of $\bar\Lambda$ from the data is ambiguous, in
the sense that it is necessarily polluted with an infrared renormalon
ambiguity and a corresponding poorly behaved perturbation series.

We can use the data (\ref{D12data}) to derive an experimental lower
bound on $\langle s_H-\overline m_D^2\rangle$.  Taking the relative
branching ratio to be 27\%, consistent with all measurements, we find
\be
  \langle s_H-\overline m_D^2\rangle\ge0.49\gev^2\,.
\ee
We can translate this into a bound on $\bar\Lambda$, which at one loop
yields
\be
  \bar\Lambda_{\rm one\ loop}
  >\left[0.33-0.07\left({\lambda_1\over0.1\gev^2}\right)\right]\gev\,.
\ee
Note that our prejudice is that $\lambda_1<0$, so it is probably
conservative to ignore the small $\lambda_1$ term. When two loop
corrections (proportional to $\beta_0\alpha_s^2$) are included, the
bound is weakened to
\be
  \bar\Lambda_{\rm two\ loop}
  >\left[0.26-0.07\left({\lambda_1\over0.1\gev^2}\right)\right]\gev\,.
\ee
The instability of these bounds when radiative corrections are
included is a direct reflection of the renormalon ambiguity.

Of more interest is the bound on $\vcb$ from the improved relation
(\ref{width2}), which has no (leading) renormalon ambiguity.
Including two loop corrections, we find ~\cite{FLS96}
\be
  \vcb>\left[0.040-0.00028\left({\lambda_1\over0.1\gev^2}\right)\right]
  \left(\tau_B\over1.60{\rm ps}\right)^{-1/2}\,.
\ee
We have left explicit the dependence on the lifetime $\tau_B$ of the
$B$ meson.  The contribution of the two loop correction to this bound
is 0.002, well within reason.  If, to be conservative, we take
$\langle s_H-\overline m_D^2\rangle=20\%$, then the bound becomes
$\vcb>0.038$.

{}From this point of view, of course, the ideal experiment would measure
$\langle s_H-\overline m_D^2\rangle$ directly, as well as higher
moments such as $\langle (s_H-\overline m_D^2)^2\rangle$.  Such a
program could lead to the best possible measurement of $\vcb$, with
theoretical uncertainties at the level of a few percent.

\section{Conclusions}

We have seen that excited heavy hadrons have a lot to teach us about
both QCD and physics at short distances.  The phenomenology of these
hadrons is extremely rich.  We have illustrated their potential by
discussing their spectroscopy, strong decays and production in
fragmentation and semileptonic decay, but by no means need this
exhaust the possibilities.  Dedicated theoretical and experimental
study of these states will pay real physics dividends in the upcoming Factory
Era.

\section*{Acknowledgements}

It is a great pleasure to thank the organizers of this Johns Hopkins
Workshop for a stimulating conference and their warm hospitality.
This work was supported by  the National Science Foundation under
Grant No.~PHY-9404057 and National Young Investigator Award
No.~PHY-9457916; by the Department of Energy under Outstanding
Junior Investigator Award No.~DE-FG02-94ER40869; and by the Alfred
P.~Sloan Foundation.

\end{document}